\documentstyle[11pt,newpasp,twoside,psfig]{article}
\begin{document}

\title{{\it Chandra} Grating Observations of AGN }
\author{Tahir Yaqoob $^{1,2}$ \& Urmila Padmanabhan $^{1}$}
\affil{$^{1}$ Johns Hopkins University, 3400 N. Charles St., Baltimore, MD21218. \\ 
$^{2}$ NASA/GSFC, Code 662, Greenbelt Rd., Greenbelt, MD20771.}  

\setcounter{page}{1}
\index{Yaqoob, T.}
\index{Padmanabhan, U.}

\begin{abstract}
The highest spectral resolution data for the 
Fe-K lines in type~I AGN, as observed with the {\it Chandra}
High Energy Grating (HEG), reveal a variety of line shapes.
However, the energies of the most prominent peak 
are all clustered tightly
around 6.4 keV (weighted mean $6.403 \pm 0.062$ keV).
If {\it all} the peaks were part of single, relativistically
broadened disk line, this would require unrealistically
fine tuning. Thus, some of the cores must originate
in distant matter (e.g. BLR, NLR, torus). 
On the other hand, in at least two AGN, the emission at 6.4 keV has been seen
to vanish on short timescales, indicating an origin close
to the central engine. For one of these (Mrk~509)
this is puzzling because the HEG and
simultaneous {\it RXTE} time-averaged spectra indicate only
a narrow line was present at that time.
Simultaneous HEG/{\it RXTE} observations for NGC~4593
indicate a broad and narrow Fe-K complex, all originating
in neutral Fe, and for F9 show a narrow, neutral Fe-K
component plus a broad, He-like component. The latter
signature has been observed in three other AGN by {\it XMM}
and may be quite common.
\end{abstract}

\vspace{-0.5cm}
\section{Introduction}

We discuss here results on the Fe-K emission lines
in type~I AGN using {\it Chandra} HEG and {\it RXTE}.   
The HEG currently provides the best spectral
resolution available in the Fe-K region
($\sim 1860 \ \rm \ km  \ s^{-1}$ FWHM at 6.4 keV).
We do not discuss the soft X-ray spectra (see McKernan \& Yaqoob 2003,
for a recent review, and references therein).
Padmanabhan \& Yaqoob (2003; hereafter PY03) 
present
the most precise measurements of the centroid energy, width, and equivalent
width of the {\it cores} of the Fe-K lines 
in a sample of type~I AGN 
using the HEG. In this paper
we extend some of those results using {\it RXTE}.

\section{The Fe-K Lines: Origin of the Core Emission}

At least part of the Fe-K emission line in
some AGN is believed to originate in a
relativistic accretion disk around the black hole (e.g. see
review by Fabian et al. 2000). However the Fe-K line profile
is in general complex (e.g. Yaqoob et al. 2001).
The total line emission may contain
a component originating in matter located far from the black hole,
such as the BLR, a
putative obscuring torus, or the NLR.
 
Preliminary results from studying the Fe-K lines
with the {\it Chandra} HEG in
a sample of type~I AGN have been given in Yaqoob et al. (2001) and
PY03.
Studies of additional, individual
objects have been presented by Fang et al. (2002),
Kaspi et al. (2002), Lee et al. (2002), and Turner et al. (2002).
The equivalent widths (EW) of the HEG Fe-K line {\it cores},
as measured with a single Gaussian, are typically in the
range $\sim 50-200$ eV. For the sample of nine type~I AGN in PY03
(i.e. excluding the type~1.5--1.9 AGNs, NGC~4151 and MCG~$-$5$-$23$-$16),
the weighted mean centroid energy is $6.403 \pm 0.062$ keV.
Only in the cases of MCG~$-$6$-$30$-$15 (Lee et al. 2002)
and NGC~3783 (Kaspi et al. 2002), and F9 (Figure 2) is the line core resolved 
at $>99\%$ confidence by
the HEG (in addition to the intermediate Seyferts,
NGC~4151 and MCG~$-$5$-$23$-$16).
For the remaining AGN, the upper limits on the FWHM are
all different. 
In fact, for data with this kind of
spectral resolution, the actual shapes of the
line intensity versus centroid energy, and 
EW versus FWHM contours carry information about the
shape of the Fe-K line away from the peak (see PY03).
Note that just because the line core
may be resolved by the HEG, it does not necessarily
mean that the core is from distant matter.

\begin{figure}[t]
\centerline{
\psfig{figure=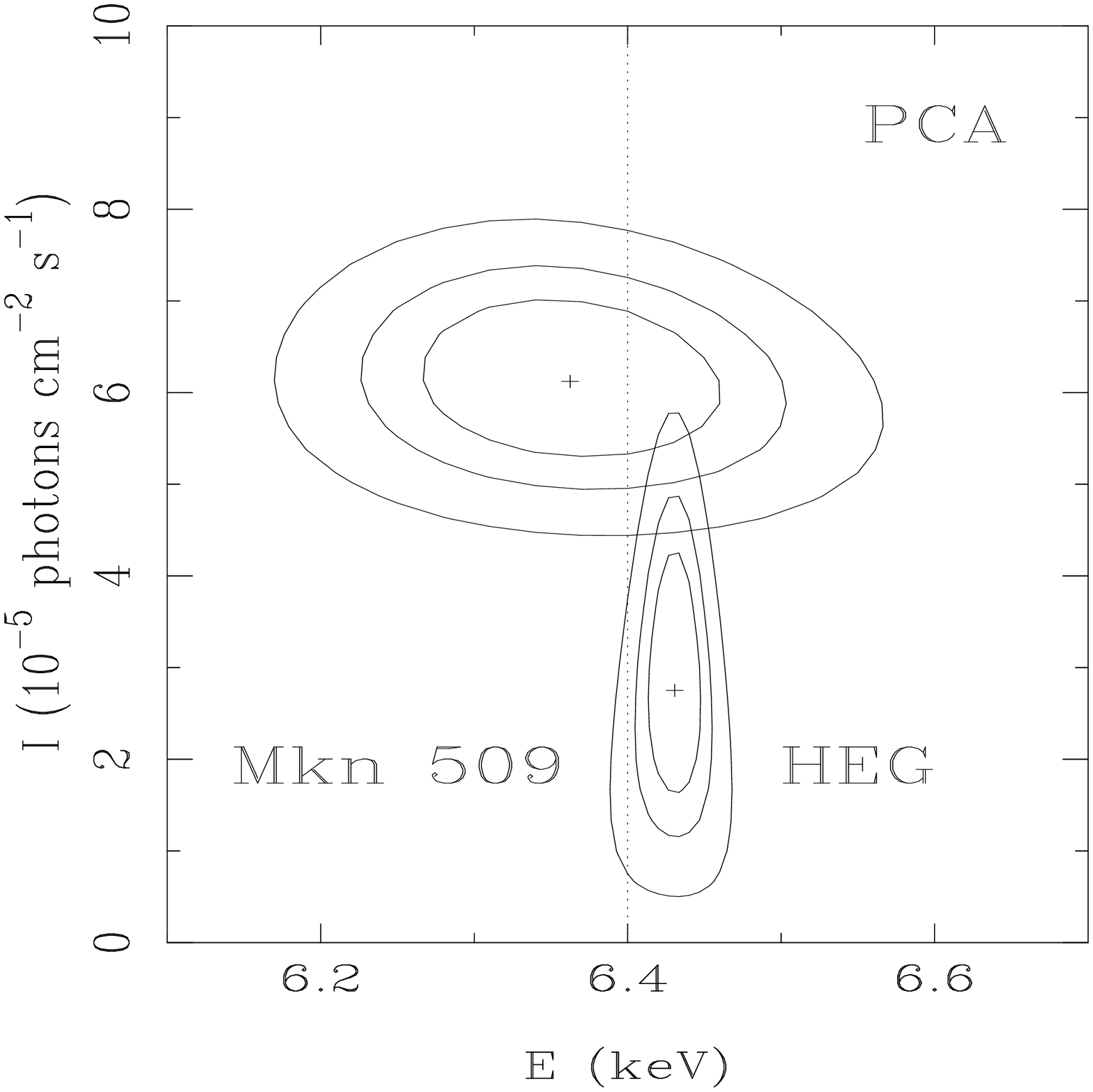,width=4.5cm,height=7cm}
\psfig{figure=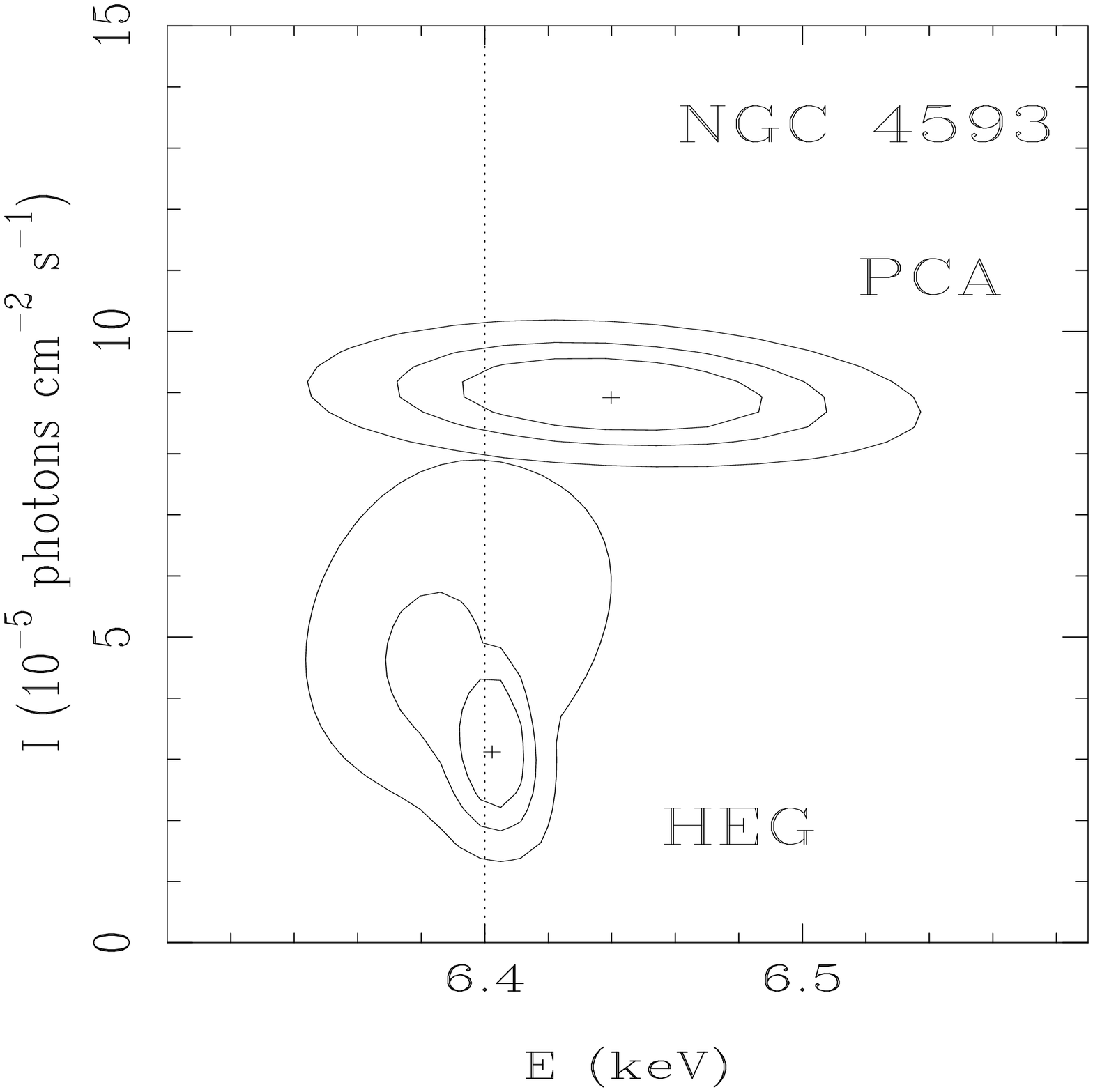,width=4.5cm,height=7cm}
\psfig{figure=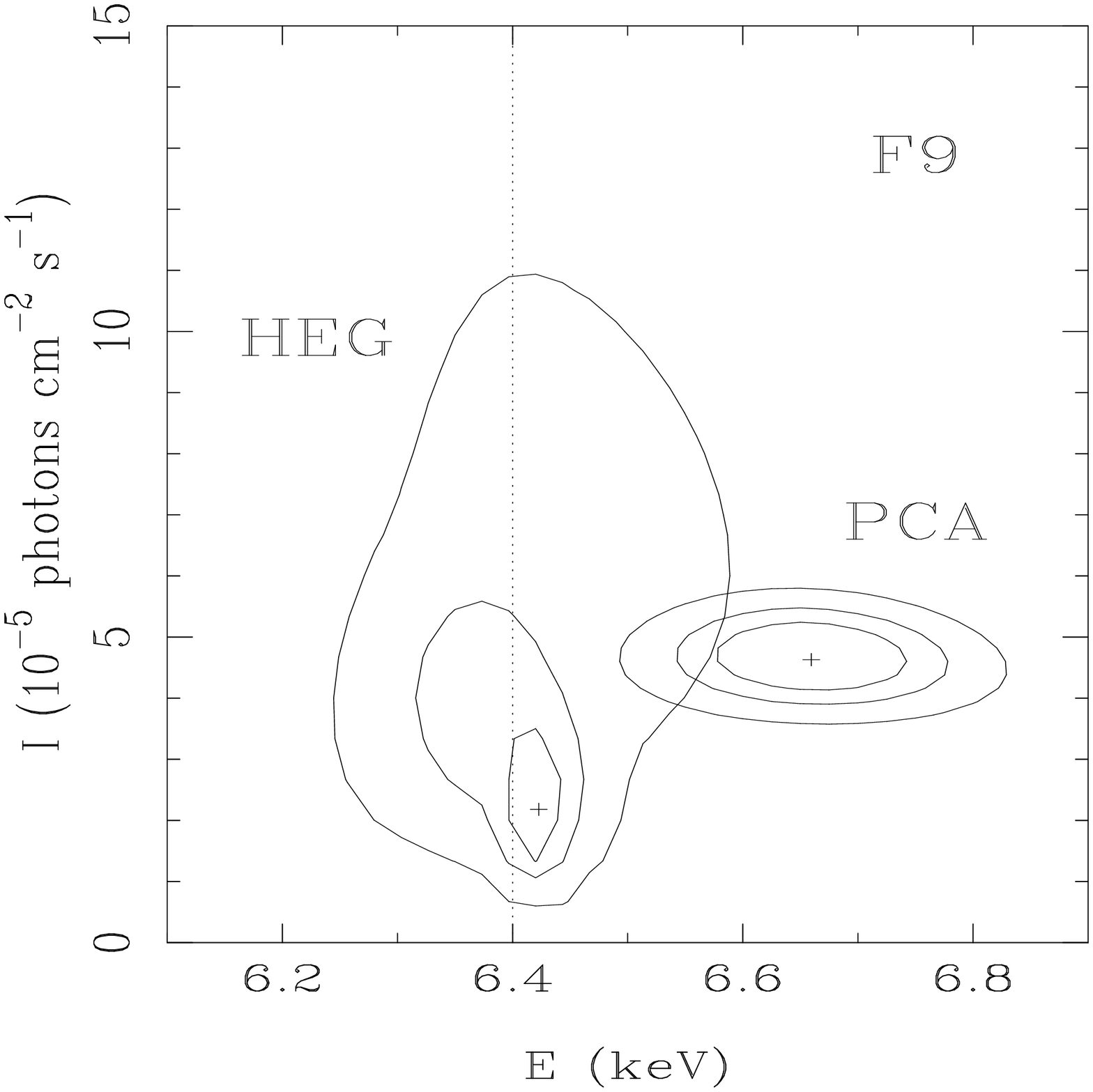,width=4.5cm,height=7cm}
}
\vspace{-2cm}
\centerline{
\psfig{figure=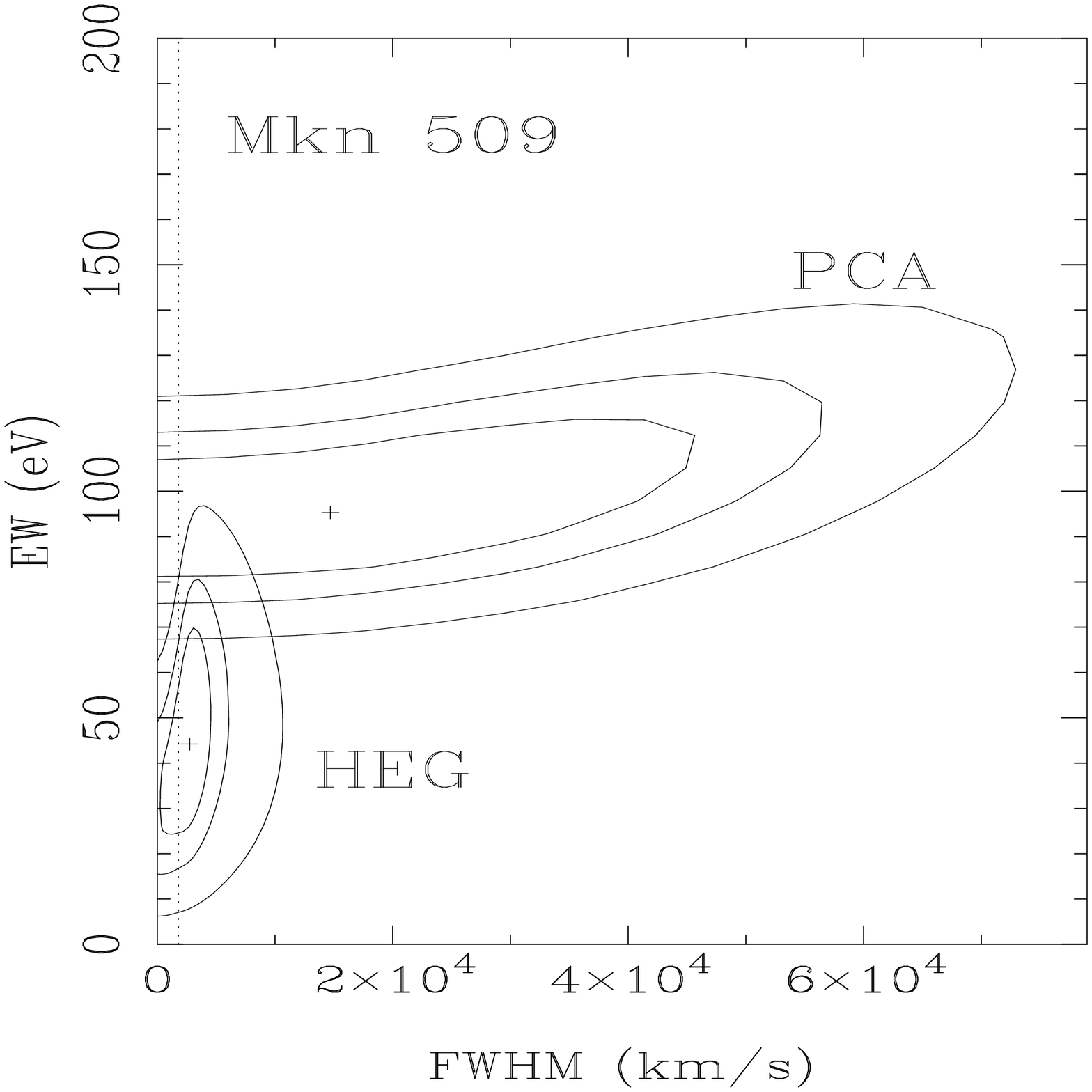,width=4.5cm,height=7cm}
\psfig{figure=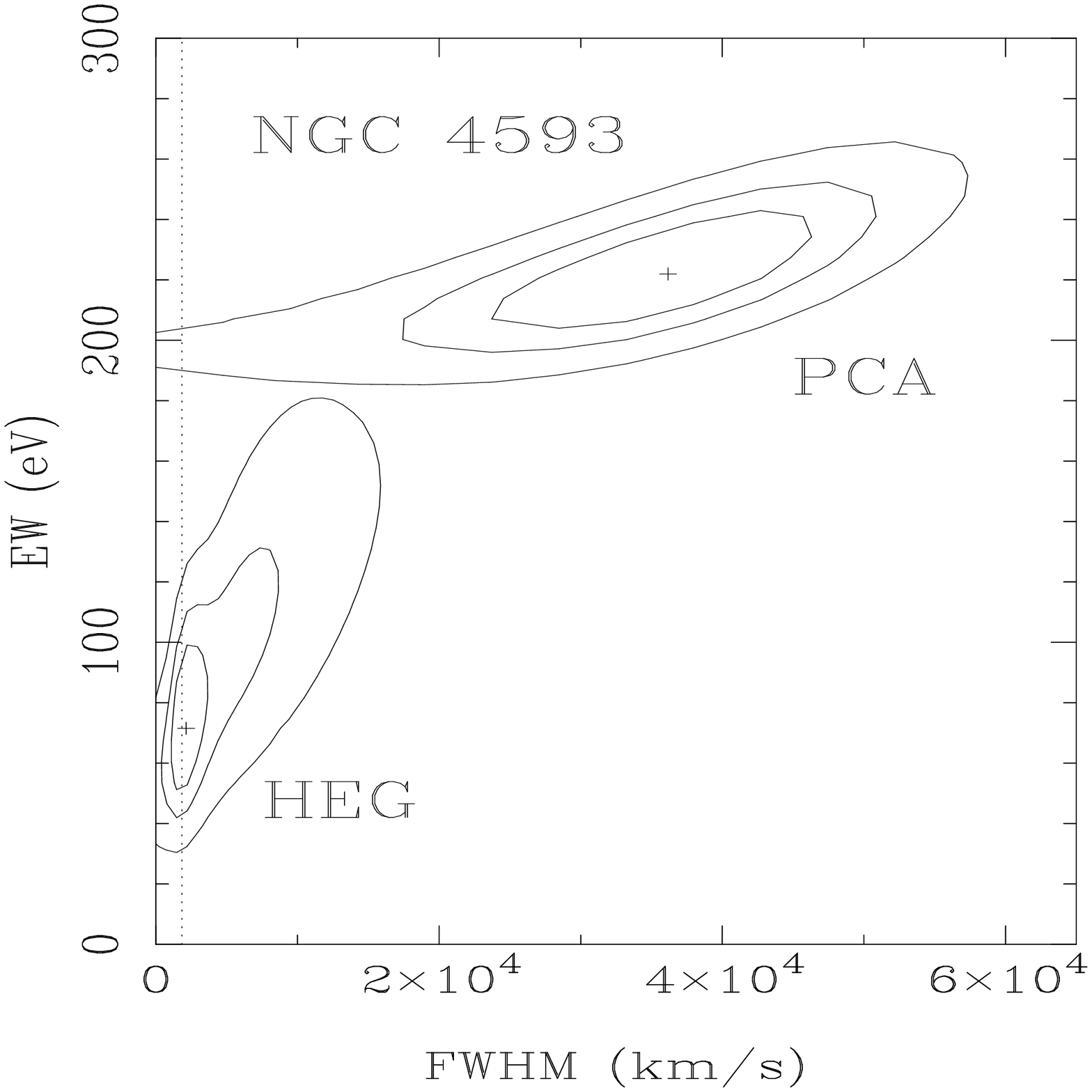,width=4.5cm,height=7cm}
\psfig{figure=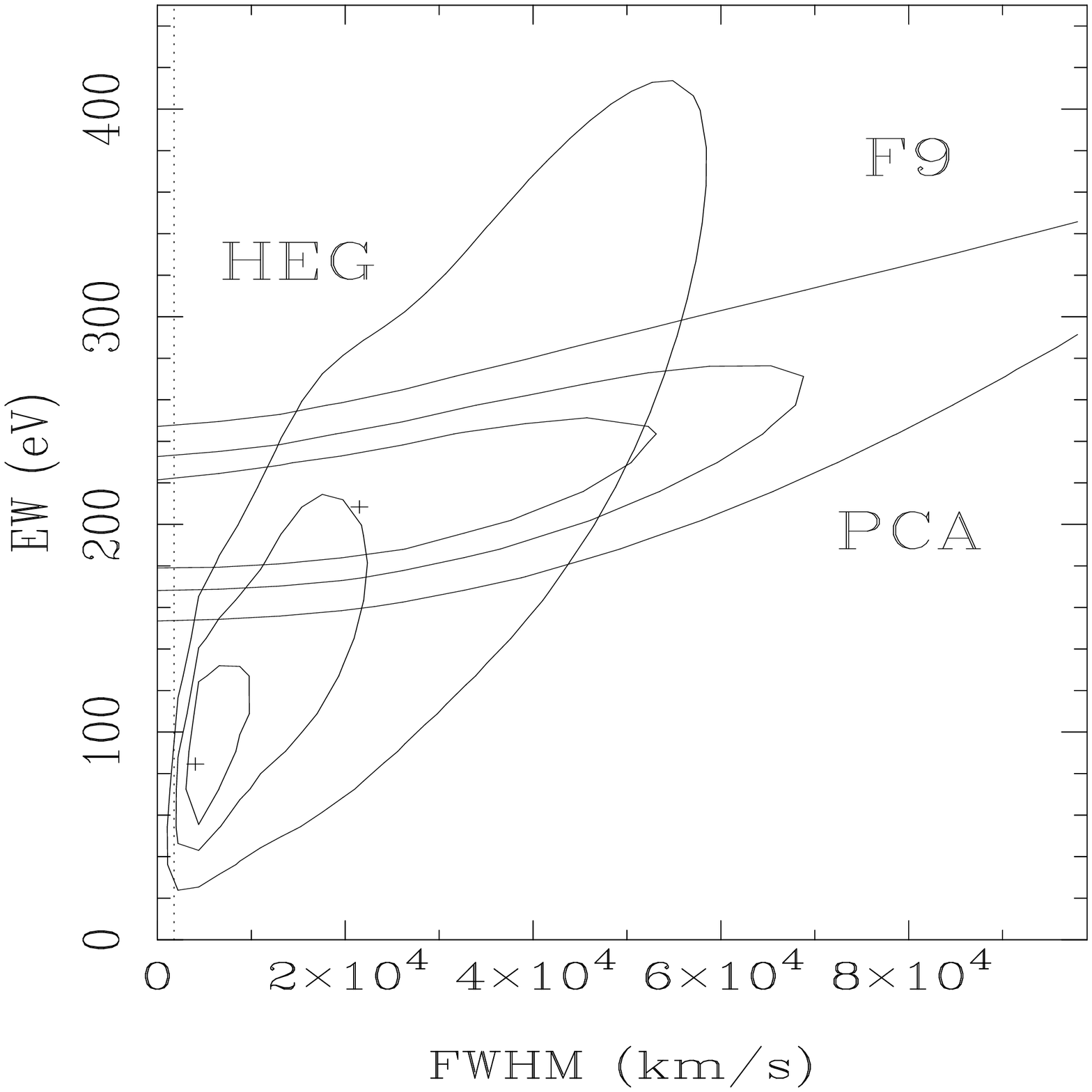,width=4.5cm,height=6.75cm}
}
\caption{HEG and PCA confidence contours (68\%, 90\%, 99\%) of
Fe-K line intensity vs. center energy (top) and EW vs. FWHM (bottom). }
\label{fig1}
\end{figure}

In some cases a broad base to the Fe-K line is clearly
detected in the {\it Chandra} data 
(e.g. NGC~4051, F9, Yaqoob et al. 2001;
MCG~$-$6$-$30$-$15, Lee et al. 2002). However, due to
the very small effective area, it is in general difficult
to detect. Either way, it is difficult from time-averaged
spectra alone, to ascertain whether the core of the Fe-K line is
from distant matter or really part of a single
broad line which has contributions from a disk
extending to large radii. Certainly, the main peak
of the line in every type~I AGN so far is at $\sim 6.4$ keV
with a very small dispersion (62 eV).
This, along with
simulations, shows that much fine tuning is
required (especially of the inclination angle)
if the line is only from a disk (see PY03).
Thus, line cores measured by {\it Chandra}
cannot {\it all} be due to the peak of a disk line.
In one case (MCG~$-$6$-$30$-$15, Lee et al. 2002), it
has been possible to deduce, from variability, that the core of the Fe-K
line is {\it not} from distant matter. However, most AGN do not
vary as rapidly as MCG~$-$6$-$30$-$15 in a single
observation so we must await further monitoring data
to establish origin of the Fe-K line core in more AGN.

\section{The Broad Fe-K Line Components}
Now we discuss some
of the {\it Chandra} observations which were simultaneous with
{\it RXTE} (Mrk~509, NGC~4593, F9). 
The {\it RXTE} PCA has more than 100 times
the effective area of the HEG in the Fe-K band. Although
the velocity resolution is $\sim 55,000 \ \rm km \ s^{-1}$
FWHM, the PCA is excellent for studying the broad component
of the Fe-K lines. 
If we use a simple Gaussian (with intrinsic
width free) to model the line in 
separate instruments, the PCA will `pick-up' more line emission than
the HEG (with possibly a different centroid energy), if the line is complex. 

Details of the HEG Gaussian fitting are described in (PY03). 
For the PCA data, in addition to a power-law
continuum and Gaussian line, we included a Compton
reflection continuum
(from neutral, optically-thick matter) with the `reflection fraction' ($R$) free,
utilizing data out to $\sim 14-15$ keV.
Fig. 1 shows the 68\%, 90\%, and 99\% confidence
contours of line intensity versus centroid energy, and 
EW versus FWHM respectively. 
We see that for Mrk~509 the HEG and PCA give consistent
results, both giving peaks at $\sim 6.4$ keV, with
PCA giving only a marginally larger EW than the HEG.
Therefore, during this observation of Mrk~509, the Fe-K
line was dominated by a narrow, unresolved (by HEG)
component with only a weak broad component.
However, {\it ASCA} monitoring data of Mrk~509 showed
that the peak at 6.4 keV can disappear and move
to lower or higher energy, {\it on a timescale of days}.
This is puzzling and would suggest that the peak of the 
Fe-K line is {\it not} from distant matter (i.e. it
is from well inside the BLR).

Next, the contours for NGC~4593 show that the HEG and PCA
line intensities and EWs {\it do not overlap} at 99\%
confidence (Fig. 1). Therefore the Fe-K line is definitely complex, with the
HEG only picking up the core, which is about half of the
intensity of the total line emission. The centroids of the
HEG core and the total are both consistent with 6.4 keV,
indicative of an origin in cold Fe for the whole line.
Variability information is required to determine whether
there are two separate line components or a single broad
line. 

The contours for F9 show that the HEG and PCA measure
lines with {\it different} centroid energies (Fig. 1). We know from
Mrk~509 and NGC~4593 that this is not a cross-calibration issue.
In F9 the HEG line peaks at $\sim 6.4$ keV and the PCA line peaks
at $\sim 6.66$ keV. So again, the line is definitely complex.
The PCA line likely includes a contribution
from He-like Fe in addition to the 6.4 keV component measured
by the HEG. This time there is not a clear distinction between
the HEG and PCA EW versus FWHM contours, because, as is
evident
from the line profile (see Yaqoob et al. 2001), the core of the
line is not as `peaky' as that in some of the other AGN
(such as Mrk~509 and NGC~4593). In fact, Fig. 1 shows
that the HEG contours 
from single-Gaussian fits pick up the whole
broad line, and indeed F9 is one of the objects in which the
broad profile is detected with the HEG. Note that a neutral,
narrow Fe-K line plus a broad He-like line has been observed
in {\it XMM} observations in at least three other AGN (Mrk~509, 
Mrk~205, Reeves, these proceedings; NGC~5506, Matt et al. 2001)
and may be a common signature.

\acknowledgements

The authors gratefully acknowledge support from
NASA grants NCC-5447 (T.Y., U.P.), NAG5-10769 (T.Y.), 
and CXO grants GO1-2101X, GO1-2102X (T.Y., B.M.).
The authors are grateful to the {\it HST}, {\it Chandra} and {\it RXTE}
instrument and operations teams for making these observations
possible. We also thank B. McKernan, I. M. George,
T. J. Turner, \& K. Weaver for their contributions to this work.

\end{document}